# Synthesis of $La_{0.5}Ca_{0.5-x}\square_{x}MnO_{3}$ nanocrystalline manganites by sucrose assisted auto combustion route and study of their structural, magnetic and magnetocaloric properties

S. Ben Moumen[1], Y. Gagou[2], M. Chettab[2], D. Mezzane[1], M. Amjoud[1], S. Fourcade[3], L. Hajji[1], Z. Kutnjak[4], M. El Marssi[2], Y. El Amraoui[5], Y. Kopelevich[6] and Igor A. Luk'yanchuk[2,7]

[1]*LMCN, F.S.T.G. Université Cadi Ayyad, BP 549, Marrakech, Morocco*
[2]*LPMC, Université de Picardie Jules Verne, 33 rue Saint-Leu, 80039 Amiens Cédex, France*
[3]*LCMCB, Université de Bordeaux, 33600 Pessac, France*
[4]*Jozef Stefan Institute, Jamova cesta 39, 1000 Ljubljana, Slovenia*
[5]*LaMCScI, Faculté des Sciences, Université Mohammed V, Rabat, Morocco*
[6]*Instituto de Fisica Gleb Wataghin, Universidade Estadual de Campinas, Unicamp 13083-859, Campinas, Sao Paulo, Brazil*
[7]*Department of Cybersecurity and Computer Engineering, Faculty of Automation and Information Technologies, Kyiv National University of Construction and Architecture, Kyiv, Ukraine*

*Corresponding authors:*

S. Ben Moumen: said.benmoumen@yahoo.fr

Y. Gagou: yaovi.gagou@u-picardie.fr

## Abstract

Perovskite manganite $La_{0.5}Ca_{0.5-x}\square_{x}MnO_{3}$ (LCMO) nanomaterials were elaborated using the sucrose modified auto combustion method. Rietveld refinements of the X-ray diffraction patterns of the crystalline structure confirm a single-phase orthorhombic state with *Pbnm* space group (No. 62). The Ca-vacancies were voluntarily created in the LCMO structure in order to study their influence on the magnetic behaviour in the system. The magnetic susceptibility was found to be highly enhanced in the sample with Ca-vacancies. Paramagnetic-to-ferromagnetic phase transition was evidenced in both samples around 254 K. This transition is, characterized by a drastic jump of the susceptibility in the sample with Ca-vacancies. The maximum of entropy change, observed for both compounds at magnetic field of 6T was 2.30 J $kg^{-1}K^{-1}$ and 2.70 J $kg^{-1}K^{-1}$ for the parent compound and the lacunar one respectively. The magnetocaloric adiabatic temperature change value calculated by indirect method was 5.6 K and 5.2 K for the non-lacunar and Ca-vacancy compound, respectively. The Ca-lacunar $La_{0.5}Ca_{0.5-x}\square_{x}MnO_{3}$ (x=0.05) reported in this work demonstrated overall enhancement of the magnetocaloric effect over the LCMO. The technique used to elaborate

LCMO materials was beneficial to enhance the magnetocaloric effect and magnetic behaviour. Therefore, we conclude that this less costly environmentally friendly system can be considered as more advantageous candidate for magnetic refrigeration applications then the commonly Gd-based compounds.

## 1. Introduction

Perovskite manganites attracted great interest in last decades for their application in magnetic refrigeration (MR) systems and in heat pumps. This application is based on magnetocaloric effect (MCE) that can replace the harmful refrigerant gases [1–3]. These systems can be used in low pressure devices with reduced energy consumption and low cost of the fabrication [4]. In these materials, the cooling efficiency dependents on their relative cooling power (RCP) [4,5]. In MCE the heat flow can be exchanged during suppression of applied field and vice versa in an adiabatic process [6,7]. Generally, there are two different ways to measure the adiabatic temperature change: indirect method, by measuring the field and temperature dependence of the magnetization M(T,H), and direct methods where the heat flow or MCE temperature change can directly be measured using DSC technique or thermistors [8]. Pecharsky *et al*. reported that gadolinium based materials present large magnetic entropy change ($\Delta S_M$) of 36 J kg$^{-1}$ K$^{-1}$ at TC = 272 K for a field change of 0–5 T [9]. According to the Maxwell thermodynamic relationship [10] and the principles of thermodynamics, the entropy variation and adiabatic temperature changes in the material can be determined by the equations:

$$\Delta S_M(\Delta H) = \int_{H_1}^{H_2} \left(\frac{\partial M}{\partial T}\right)_H dH, \tag{1}$$

and

$$\Delta T_{MCE} = - \int_{H_1}^{H_2} \frac{T}{C_P} \left(\frac{\partial M}{\partial T}\right)_H dH \tag{2}$$

respectively [11]. Here $\Delta S_M$ is the magnetic entropy change, $H_2$-$H_1$ is the change of the applied field, M is the magnetization, H the applied magnetic field and T is the current temperature. The parameters Cp and $\Delta T$ are the specific heat of the sample and the magnetocaloric temperature change in an adiabatic process, respectively. Nowadays, many researches are focused on the magnetic materials exhibiting large magnetocaloric effect near room temperature for their application as a refrigerant material in magnetic cooling technologies [12,13]. Consequently, the ferromagnetic perovskite material, LaMnO$_3$, has attracted the attention of researchers as perspective magnetocaloric material [14,15]. Furthermore, the creation of lacuna [16] or the insertion of cations such as Bi, Ca and Sr in A or B site of the perovskite separately or simultaneously [17,18] lead to the modification of such physical properties [19] as charge-ordering [20] and orbital density [21]. In general, the valence of Mn ions ($Mn^{3+}$ vs. $Mn^{4+}$) controls the magnetic and electrical behaviour in these compounds. This material family had been extensively investigated because of its numerous potential applications [22]. Their structural properties and their remarkable sensitivity to Jahn-Teller distortions [23] are closely dependent on the preparation routes. Magnetic and electrical transitions as well as colossal magnetoresistance (CMR) have been studied previously in this system. In addition, their MCE gains interest due to their near room temperature large entropy change, 8.3 J $kg^{-1}$ $K^{-1}$ at 270 K under a magnetic field of 5 T, as reported by Andrade et *al* [24]. Moreover the manganites seem to be very promising MCE candidates, since their transition temperature and the magnitude of their $\Delta S_M$ can be strongly modified by adjusting their chemical composition [25] and their microstructure according to the elaboration method and its conditions [26]. Many preparation methods are used to synthesize these manganites including solid state reaction [27], ball milling technique [28], sol-gel Pechini method [16], and hydrothermal method [29,30]. Likewise, pure and alkaline-earth-substituted lanthanum manganites have been synthesized by self-propagating high-temperature synthesis (SHS) that is a low-cost and attractive method for rapidly generating large amount of nanosized manganites perovskites [31]. Urea and glycine are the commonly used as fuels in these combustion synthesis [32]. Since the magnetocaloric properties of manganites were investigated only in a limited extent, the search for other compositions and new preparation methods which may lead to an enhancement of the magnetocaloric effect is still under the initial stage. In present paper we present the novel sucrose assisted auto combustion method to prepare $La_{0.5}Ca_{0.5-x}\square_{x}MnO_3$ nanocrystalline manganites. This elaboration technique is used in order to investigate the correlation between microstructural, magnetic, and magnetocaloric properties of the synthesized $La_{0.5}Ca_{0.5-x}\square_{x}MnO_3$ system. The results show an enhancement of magnetocaloric properties and the amplification of the magnetization in the Ca-vacancy compound.

## 2. Experimental

### 2.1 Synthesis

Single phase polycrystalline manganites nanopowders $La_{0.5}Ca_{0.5-x}\square_{x}MnO_3$ (LCMO) with x=0 and 0.05, have been prepared using the sucrose assisted auto combustion method. For each sample, stoichiometric amounts of $La(NO_3)_3 \cdot 6H_2O$, $Ca(NO_3)_2 \cdot 4H_2O$ and $Mn(NO_3)_2 \cdot 4H_2O$ were mixed and added at the room temperature to an appropriate amount of sucrose $C_{12}H_{22}O_{11}$ used as fuel. The mixture was heated in a sand bath at 120 °C for 2 hours leading to the swelling of the powders due to the evaporation of the gases generated during the reaction which produced a foamy mass. Right after, a spark appeared spontaneously and propagated on the surface of this foamy mass giving rise to a very fine powder (Fig.1). Then this "as-prepared" powder was heat-treated at 850°C for 8h in air using a controlled heating rate of 2°C/min. Finally, the nanosized smooth oxide black powders were obtained. The elaborated samples were denoted S0 and S5 referring to the calcium lacuna ratio of 0 and 5% respectively. Ca-vacancies are created in $La_{0.5}Ca_{0.5-x}\square_{x}MnO_3$ sample (S5) by decreasing the stoichiometric amount of the calcium nitrate $Ca(NO_3)_2.4H_2O$ precursor used during the preparation. All the reagents, purchased from Sigma-Aldrich, have analytical grade and were used without any further purification. The crystallographic structure of the samples was analyzed at room temperature. The X-ray patterns were recorded using the diffractometer X'Pert PRO with Cu-Kα (wavelength λ=1.5405 Å) radiation operating at 40 kV and 40 mA. The patterns were recorded in a range from 10° to 80° with the step size of 0.017°. The cell parameters, atomic positions and thermal agitation factors were refined satisfactorily from diffraction patterns by using Rietveld method via Fullprof software [4]. TGA measurements were performed by using SDT Q600 thermal analyzer in air with a heating rate of 5 °C/min from room temperature to 1000 °C. Samples grains morphology was observed by using a high resolution JEOL Field Emission Gun-Scanning Electron Microscope (FEG-SEM). Magnetic properties were obtained by using a Physical Property Measurement System (PPMS-DynaCool) Quantum Design apparatus operating in temperature range 2 - 300K and in magnetic field range 0 - 9T. The magnetization was measured by using vibrating sample magnetometer (VSM) method, integrated in the system.

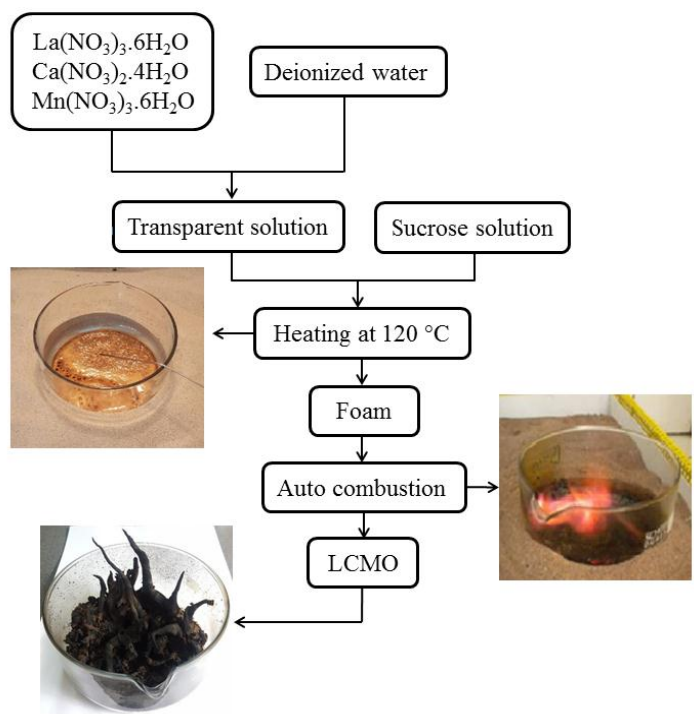

Fig. 1. Flow chart for the preparation of LCMO

## 3. Results and discussion

### 3.1 Thermal behavior

Thermo-gravimetric analysis of the as-prepared powders is depicted in Fig. 2. Three different regions of weight loss were observed when increasing temperature. The first weight loss of about 2.74 % in the range from ambient to 343 °C can be attributed, for both prepared powders, to the evaporation of physisorbed coordinated water and probably to the decomposition of the nitrates [33]. Subsequently, the second weight losses of 11.84% between 343 and 646 °C can be ascribed to the complete decomposition of the remaining sucrose, in non-lacunary and calcium-lacunary calcium powders. The last weight losses of 1.63 % situated in the range from 655 to 1000 °C can be attributed to the progressive powders oxidation of the samples. It is worth to mention that the total oxidation of Ca-deficient oxide occurs at ~ 647 °C that is lower than that of non-lacunary oxide (~ 656 °C). Besides, the offset observed between the curves at high temperatures can be considered as a confirmation of the creation of the Ca-lacuna in S5 sample.

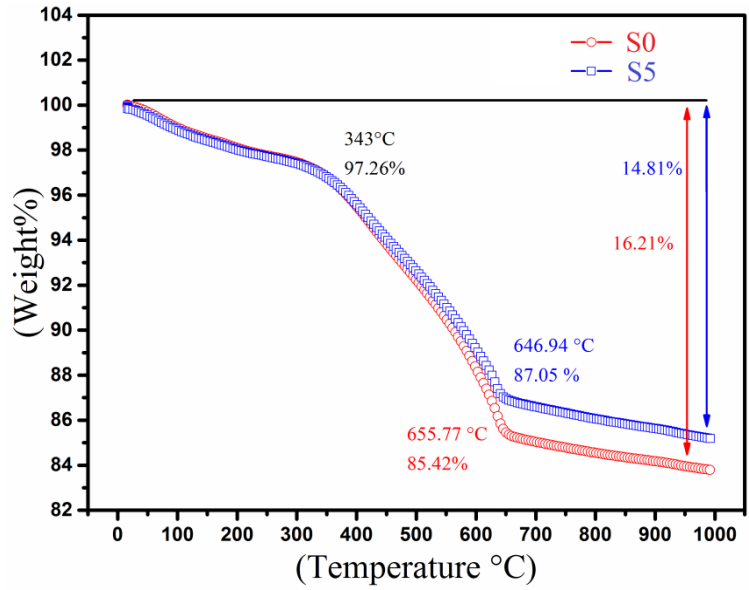

Fig. 2. TGA profiles of as-prepared powders in air with heating rate of 5 °C/min.

### 3.2 Chemical analysis

The stoichiometry of the manganese and oxygen ions present in our samples was checked using the oxidation reduction method as described in refs [34–38]. According to the chemical formula $La^{3+}_{0.5}Ca^{2+}_{0.5-x}\square_{x}Mn^{3+}_{0.5-2x}Mn^{4+}_{0.5+2x}O_3$, by creating the Ca-lacuna, we will favour the increase of $Mn^{4+}$ ions in our sample. The powder of the samples is dissolved in a mixture of sulfuric acid ($H_2SO_4$) and oxalic acid ($H_2C_2O_4$) solution and heated at 50 °C until total dissolution. A reaction occurs between the manganite and the oxalic acid which led to total reduction of both $Mn^{3+}$ and $Mn^{4+}$ into $Mn^{2+}$ and the excess of the oxalic acid was titrated using 0.0200±0.0003M potassium permanganate ($KMnO_4$) solution under stirring. The proportions of $Mn^{3+}$ and $Mn^{4+}$ and O ions stoichiometry were analysed quantitatively and gathered in Table. 1. As it can be seen, the experimental results are in good agreement with the theoretical data confirming the amount of deficiency in the calcium-vacancy sample (S5).

Table. 1. Chemical analysis results for samples S0 and S5.

|  | Theoretical (%) | | | Experimental (%) | | |
| --- | --- | --- | --- | --- | --- | --- |
|  | $Mn^{3+}$ | $Mn^{4+}$ | O | $Mn^{3+}$ | $Mn^{4+}$ | O |
| (S0) $La_{0.5}Ca_{0.5}MnO_3$ | 50 | 50 | 3 | 49.68±1.07 | 50.32± 1.52 | 3.05±0.07 |
| (S5) $La_{0.5}Ca_{0.45}\square_{0.05}MnO_3$ | 40 | 60 | 3 | 41.55±0.88 | 58.45± 1.75 | 3.09±0.07 |

### 3.3. Structural study

Fig. 3(a) and 3(b) show the powder X-ray diffraction patterns of the samples S0 and S5, respectively, recorded at room temperature. The observed peaks have the sharp well-defined profiles that bear out the good crystallinity of both calcined powders. All the diffraction lines have been indexed in the orthorhombic structure symmetry with *Pbnm* (No. 62) space group. These diagrams showed well defined perovskite structure without any impurity phases. The Ca-deficiency did not modify the parent compound structure. The inset of Fig.3 shows the crystal structure of the parent compound designed using the software of visualization for electronic and structural analysis (VESTA) [39]. The structural refinement was performed assuming an orthorhombic perovskite structure with *Pbnm* space group where Lanthanum, Calcium and Manganese ions occupy the 4c(x,y,1/4) and 4b(0,1/2,0) special positions, respectively, with multiplicity 1, while the Oxygen anions occupy general position 8d(x,y,z) and 4c with multiplicity 1 and 2, respectively. The profile fitting was started with the scale, zero point and background parameters followed by the unit cell parameters led to rapid convergence. A pseudo-Voigt function was used to describe the individual line profiles, the peak asymmetry and preferred orientation corrections were applied. Finally, the atomic position parameters and the individual isotropic thermal agitation parameters were refined. Patterns of the experimental, calculated diagrams and their differences were plotted in Fig.3 (a) and (b) for sample S0 and S5, respectively. Their refined cell parameters, unit cell volume, atomic positions and reliability factors were gathered in Table 2. The small difference of relative intensity between calculated and experimental patterns and ($\chi^2$) confirms the validity of refinement. Previous study reported by Gonzalez-Calbet *et al.* on $La_{0.5}Ca_{0.5}MnO_3$ samples showed that the different Oxygen stoichiometry induced splitting of some structural lines that are clearly observable in the X-ray patterns. Thus, the extra planes (111), (121) and (311) reflections observed mainly in X-ray diagrams of S0 and S5, marked by symbol #, reveal the absence of oxygen vacancies, with content very close to 3 [6]. This observation corroborates the results found previously by the chemical analysis (Table 1).

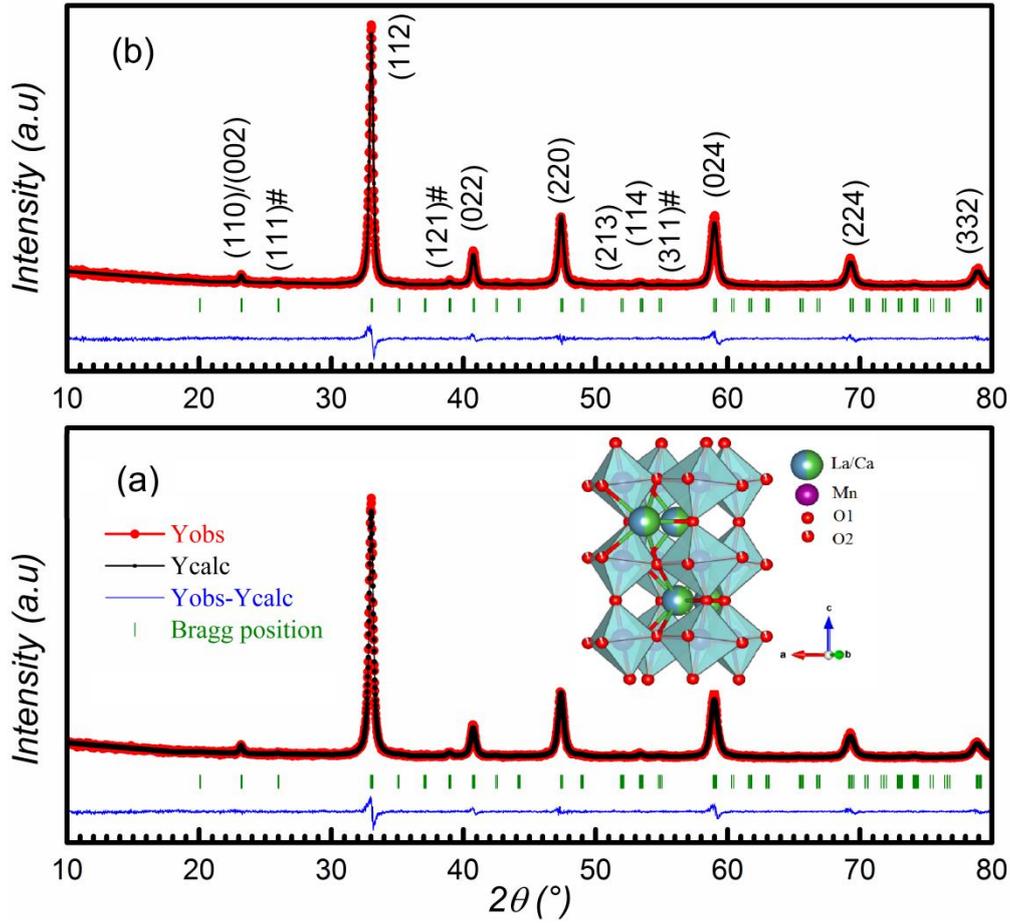

Fig.3. The Rietveld refinement parameters deduced from X-ray diffraction patterns at room temperature (a): S0 and (b): S5. Note that the # marked (*hkl*) planes are the indication of oxygen content too close to 3, as reported in Ref [6].

Table.2. Structural parameters for S0 and S5 samples obtained from Rietveld refinement.

| | x | y | z | Occ |
|---|---|---|---|---|
| $La_{0.5}Ca_{0.5}MnO_3$ (S0): | | | | |
| Lattice parameters: a=5.4159 Å, b= 5.4289 Å, c= 7.6730 Å, V= 225.6077 Å$^3$. | | | | |
| Crystal system: orthorhombic. Space group: *Pbnm* | | | | |
| $R_{Bragg}$= 2.51, $\chi^2$= 1.08, Rp=18.2, Rwp=18.0 | | | | |
| La | 0.00027 | 0.01498 | 0.25000 | 0.48946 |
| Ca | 0.00027 | 0.01498 | 0.25000 | 0.50050 |
| Mn | 0.00000 | 0.50000 | 0.00000 | 1.01158 |
| $O_1$ | 0.06392 | 0.49559 | 0.25000 | 1.12307 |
| $O_2$ | 0.71073 | 0.25636 | 0.02144 | 1.78084 |

$La_{0.5}Ca_{0.45}\square_{0.05}MnO_3$ (S5) :

Lattice parameters: a=5.4276 Å, b= 5.4227 Å, c= 7.6711 Å, V= 225.7779 Å$^3$.

Crystal system: orthorhombic. Space group: *Pbnm*

$R_{Bragg}$= 3.53, $\chi^2$= 1.24, Rp=19.7, Rwp=19.7,

| | | | | |
|---|---|---|---|---|
| La | -0.00224 | 0.01634 | 0.25000 | 0.48662 |
| Ca | -0.00224 | 0.01634 | 0.25000 | 0.44618 |
| Mn | 0.00000 | 0.50000 | 0.00000 | 1.00408 |
| O$_1$ | 0.06220 | 0.50189 | 0.25000 | 1.13394 |
| O$_2$ | 0.71310 | 0.27460 | 0.01765 | 1.82382 |

As shown in the Table 2, it is clear that Ca-vacancy increases the unit cell volume. Such result has been observed in Refs [7,40]. The A site vacancy suggest a partial conversion of $Mn^{3+}$ to $Mn^{4+}$ producing an increase of the $Mn^{4+}$ content above 50%. The samples are stoichiometric in oxygen; as a result the increase of $Mn^{4+}$ in the lacunar sample is due only to the calcium deficiency. Thus, $Mn^{4+}$ ions play amphoteric role in this matrix in favor of Jahn-Teller exchanges [41]. The tetravalent manganese ion possesses the smaller ionic radius ($r(Mn^{4+})$ = 0.53 Å) compared with the trivalent one ($r(Mn^{3+})$ = 0.64 Å) [42], which induces a decrease of the B site radius. This result cannot explain the increase of the unit cell volume but consequently other parameters as the evolution of the average ionic radius $<r_{A\text{-site}}>$ of the A site induced by the lacuna may explain such behavior [43]. According to W. Boujelben et al [44] and for electrostatic considerations, the vacancy has an average radius $<r_v>$ non equal to zero. Boujelben reported in the lacunar (Pr, Sr) MnO$_3$ manganites, that $<rv>$ is smaller than that of $Sr^{2+}$ ($r(Sr^{2+})$ = 1.31Å) and larger than that of $Pr^{3+}$ ($r(Pr^{3+})$ = 1.179 Å). From these results we can conclude that that the $<r_v>$ is larger than $r(Ca^{2+})$ one (1.18 Å). Hence the calcium vacancy will enhance the average radius of the A-site $<r_{A\text{-site}}>$, which can explain the increase of the cell volum. In the perovskite orthorhombic structure the Mn atom is surrounded by six oxygen atoms forming an irregular octahedron characterized with three different Mn-O bond length distances and two Mn-O-Mn bond angles. The atomic positions of the both manganites are gathered in Table.3. These obtained results are exploited to calculate distance and displacement between neighbored atoms to understand this phenomenon. Moreover, the ratio c/a $< \sqrt{2}$ is characteristic of a cooperative Jahn-Teller deformation that induced the distortion of MnO$_6$ octahedral and the rotation from the ideal perovskite orientation. For a distorted perovskite using the average values of <Mn–O–Mn> bond angle and <Mn–O> bond length listed in Table.3, we can evaluate the bandwidth W [45] given by the equation:

$$W = \frac{\cos(\theta/2)}{<Mn-O>^{3.5}}, \text{ where } \theta = (180° - <Mn\text{-}O\text{-}Mn>). \tag{3}$$

The obtained values of $W$ are 0.0975 and 0.0972 for S0 and S5, respectively. Consequently, using the combustion synthesis method the lacunar sample S5 present the lowest value of $W$. Basing on these results, we predict that the sample S5 should have the lowest value of Curie temperature. Since an increase in the content of $Mn^{4+}$ strengthened in the lacunar sample, the super-exchange interaction $Mn^{4+}$-O-$Mn^{4+}$ occurred with the profit of the ferro-double exchange interaction $Mn^{3+}$-O-$Mn^{4+}$ [46–48].

Table. 3 Interatomic distances and angles for S0 and S5 samples.

|  | $La_{0.5}Ca_{0.5}MnO_3$ | $La_{0.5}Ca_{0.45}\square_{0.05}MnO_3$ |
|---|---|---|
| Mn–O1 (Å) | 1.94940(15) | 1.94720(16) |
| Mn–O21 (Å) | 2.0573(6) | 1.9816(1) |
| Mn–O22 (Å) | 1.8067(5) | 1.8897(11) |
| <Mn–O (Å)> | 1.9378(3) | 1.9395(0) |
| Mn–O1–Mn (°) | 159. 4617(0) | 160.0803(0) |
| Mn–O2–Mn (°) | 165.6589(0) | 163.8441(0) |
| <Mn–O–Mn> (°) | 162.5603(0) | 161.9622(0) |

### 3.4. Microstructure analysis

Typical FEG-SEM images of S0 and S5 samples, shown in Fig.4 (a) and (b) respectively, demonstrate that the powders are composed from homogenous particles of spherical shape. The average values of grain sizes are found to be around 63 and 46 nm for S0 and S5, respectively. The agglomeration degrees calculated for each sample are 2.4 and 1.4 respectively. This shows that nano-crystalline LCMO powders have been synthesized via an assisted auto-combustion route using sucrose as fuel.

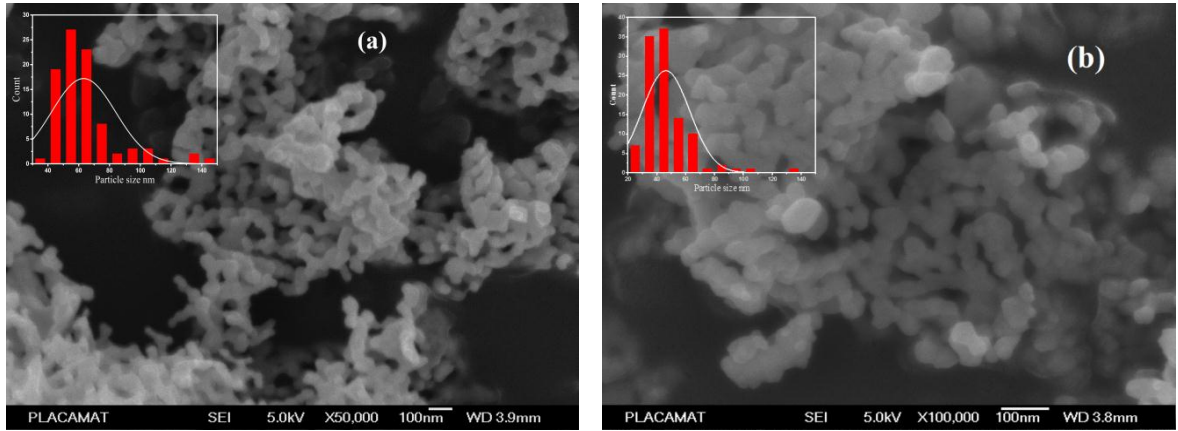

Fig.4. FEG-SEM micrographs of (a) S0 and (b) S5 nanopowders, the insets show the grain size distribution histograms adjusted to a normal law.

3.5. Magnetic study

The variation of magnetization as a function of temperature, M(T), was measured in field cooling (FC) mode under an applied field of 1T and of 0.1T(Fig.5). The observed behaviour M(T) reveals a ferromagnetic-paramagnetic phase transition in both samples. The Curie temperature Tc, determined using the derivative dM/dT, shows a transition temperature at 256 K and 254 K for S0 and S5, respectively. The small difference in the critical temperatures corroborates the obtained values of bandwidth $W$.

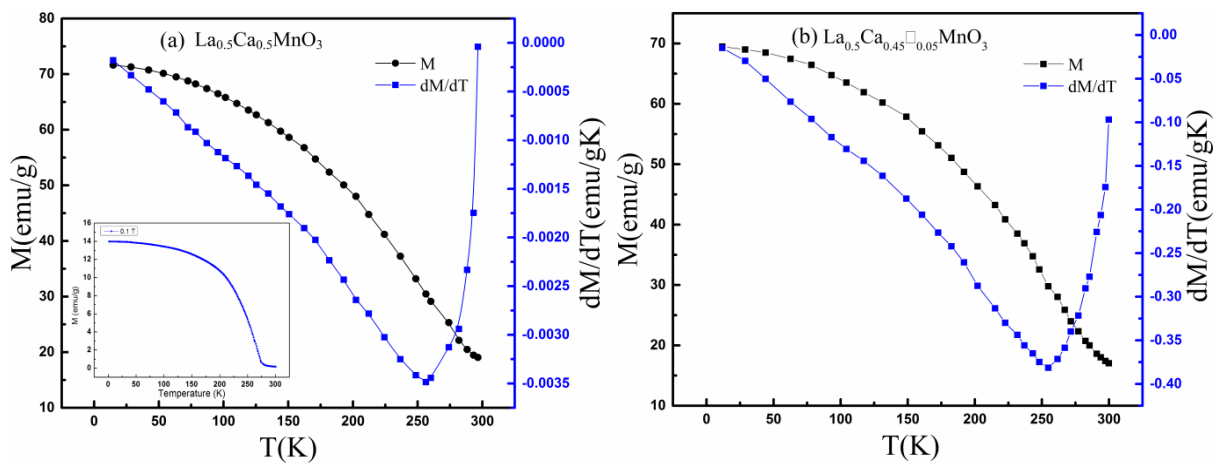

Fig.5. Temperature dependence of magnetization for (a) S0 and (b) S5 samples, the inset in Fig. 5 (a) shows the variation of magnetization as function of temperature at an applied magnetic field of 0.1T.

It is worth noting that using the sucrose combustion method, the Tc value in the stoichiometric S0 sample was observed to be higher than that measured by Walha *et al.*[49] in bulk $La_{0.5}Ca_{0.5}MnO_3$ (230 K) using the conventional solid state reaction. This result underlines the effect of the reduction of grain size using the sucrose combustion technique and inducing a strong ferromagnetic interaction in nano-particle grained samples. A similar behaviour was observed in $La_{0.7}Ca_{0.3}MnO_3$ [50]. Moreover, the antiferromagnetic interaction linked to charge ordering in $La_{0.5}Ca_{0.5}MnO_3$, as observed at low temperature TCO = 150 K in [51] is destroyed in S0. This can be explained by that, the $Mn^{4+}/Mn^{3+}$ ratio in S0 is non-equal to 1. In Fig.6, we plot the M(H) hysteresis in fields up to 6T at several representative temperatures. We observe the paramagnetic-like linear behaviour above $T_C$ and flat M(H) curve. Just below $T_C$, the magnetization increases sharply, even for field less than 0.5 T. When temperature decreases further the M(H) dependence shows the saturation above 1T, which is characteristic of a typical ferromagnetic behaviour. Therefore, the M(H) curves for both samples confirm the existence of different magnetic states as a function of temperature. The value of the spontaneous magnetization is calculated by projecting the linear part of the magnetization at high magnetic fields.

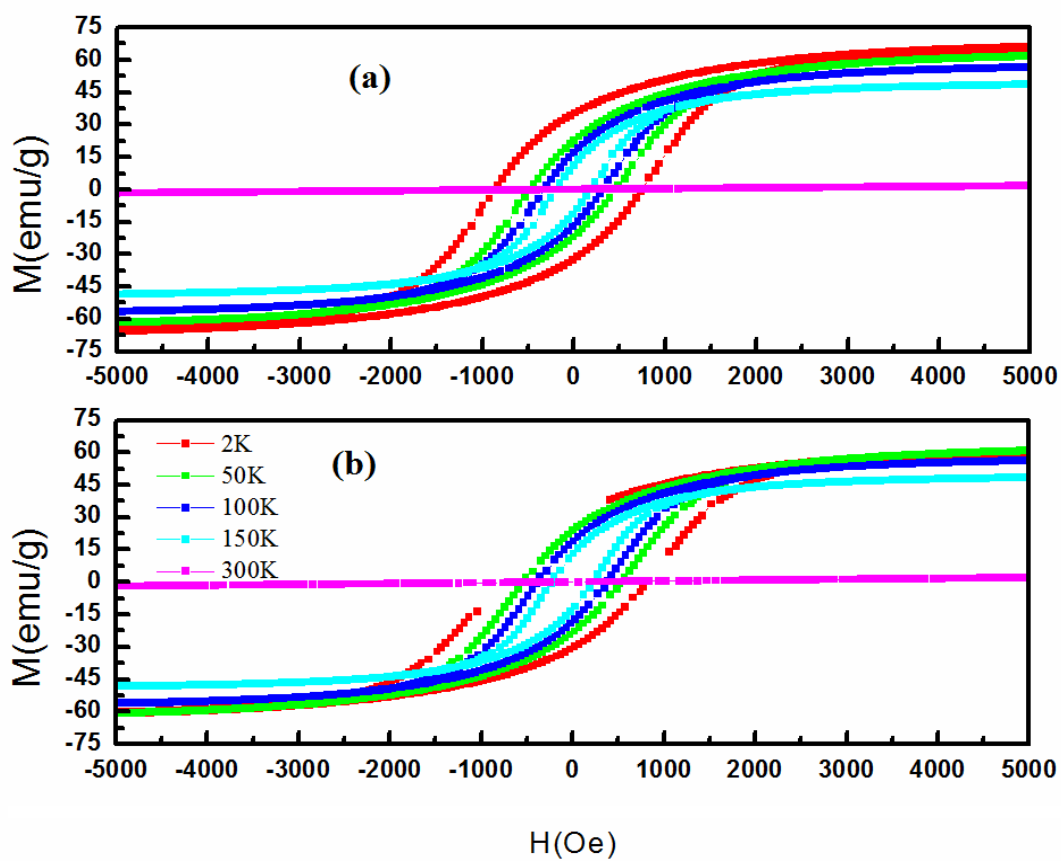

Fig.6 . Isotherm magnetization curves M(H) for (a) S0 and (b) S5 samples.

At 2K, the values of the total spins were calculated by considering of $Mn^{3+}(t^3_{2g}eg^1: S=4/2)$ and $Mn^{4+}(t^3_{2g}eg^0: S=3/2)$ configurations. The spontaneous magnetization for both samples was expressed by:

$$M_{sp} = 2[4/2 \times (0.5-2x) + 3/2 \times (0.5+2x)] \; \mu_B/Mn, \qquad (4)$$

where x is the content of calcium vacancies and $\mu_B$ is the Bohr magneton. For S0 and S5 samples the measured spontaneous magnetizations at T = 2K are reached at around 71.5 emu/g (~ 3.04 $\mu_B$/Mn) and 68.3 emu/g (~ 2.89 $\mu_B$/Mn), respectively while the calculated values for full spin alignment are 3.5 $\mu_B$/Mn and 3.4 $\mu_B$/Mn, respectively. The deviation between the experimental and theoretical values could be explained by the non-full alignment of the spin (canted state) at low temperature observed also in several other studies [52,53]. In a similar way we notice that remnant magnetization (Mr) decreases from 34.95emu/g in S0 to 30.54 emu/g in S5 while the coercive field (Hc) is observed to increase from 751 Oe in S0 to 812 Oe in S5. While introducing the Ca-vacancies, we noticed that the value of the magnetization decreases. This can be explained by the decrease of high spin ions $Mn^{3+}$, concentration and the increase of the low spin ions $Mn^{4+}$, concentration. The same behavior was reported in $La_{0.8}Ca_{0.2-x}\square_xMnO_3$ [54].

### 3.6. Magnetocaloric study

Magnetocaloric effect is investigated in S0 and S5 samples by indirect method based on the measured M(T,H) curves. Polynomial fits were performed on the upper branches of all the measured M(H) curves at all experimental temperatures. This permits us to calculate for each applied magnetic field between 0 to 6T, the evolution of magnetization as a function of temperature or for each temperature the evolution of magnetization as a function of the magnetic field. The satisfactory obtained polynomial degree was 27. Thus, the pyromagnetic coefficient $\left(\frac{\partial M}{\partial T}\right)_H$ had been easily calculated. Entropy and the adiabatic variation of the magnetocaloric temperature have been calculated according to the equations (1) and (2). Fig.7 shows the upper branches of the evolution of magnetization as a function of the temperature. Both the experimental and fitted curves are presented, demonstrating successful fits. It should be noted that the magnetization decreases with increasing of the temperature.

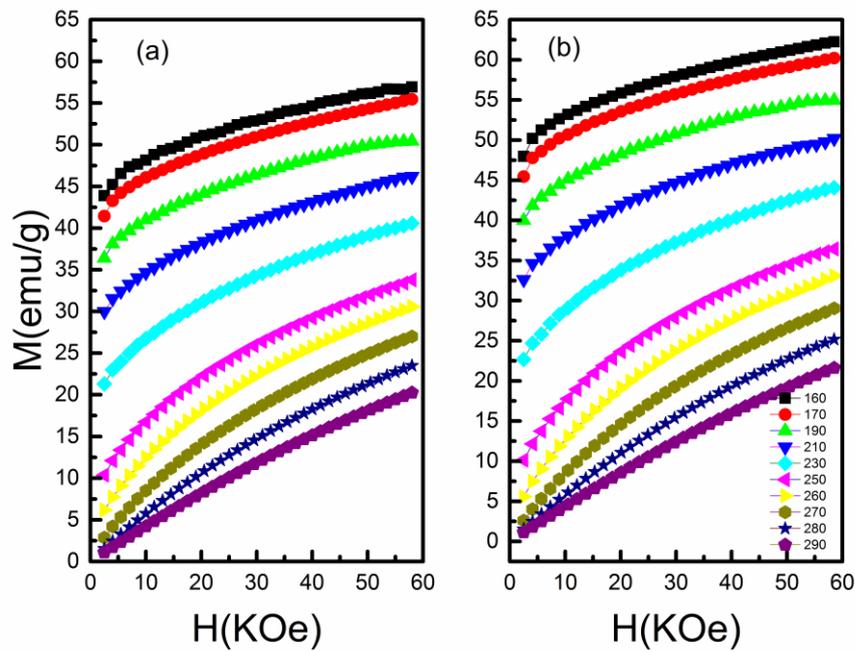

Fig.7 Variation of the upper branches of magnetization as a function of magnetic field for several temperatures in (a) S0 and (b) S5.

From M(T) curve fitting the evolution of magnetization as function of temperature was determined for several values of the magnetic field. The results were plotted in Fig.8. These curves highlight the ferromagnetic-paramagnetic phase transition in accordance with M(T) experimental results presented in Fig.5. Saturated magnetization seems to be strengthened in S5 under 6T at 150K and its value is of about 64,5 emu/g.

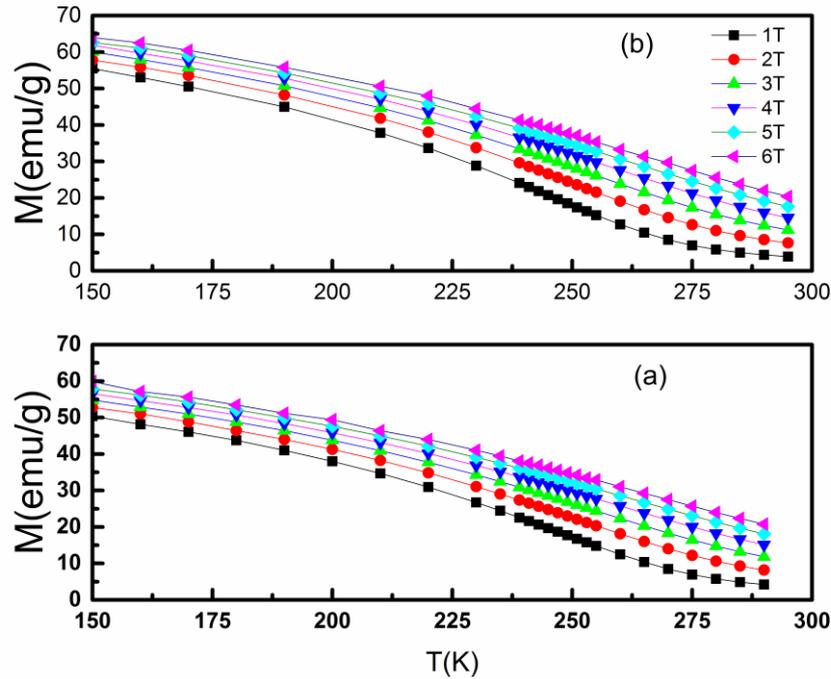

Fig.8. Variation of magnetization as a function of temperature for several values of applied magnetic field in (a) S0 and (b) S5.

Generally, the magnetocaloric properties are amplified around the Curie temperature due to the phase transition mechanisms. This amplification is also related to the order and nature of the phase transition. It should be noted that the first order transition contributes to higher variation of entropy which is related to the latent heat variation and consequently higher pyromagnetic coefficient $\left(\frac{\partial M}{\partial T}\right)_H$. The entropy was calculated for both samples and plotted in Fig.9. The entropy is maximized at $T_C$ and increases with application of the magnetic field. The corresponding temperature of the maximum evolved slighly toward the higher temperature in accordance with the theoretical predictions. The entropy value is higher in the S5 sample in comparison with the S0 sample. This can be explained by spin reordering in the Ca-vacancy matrix.

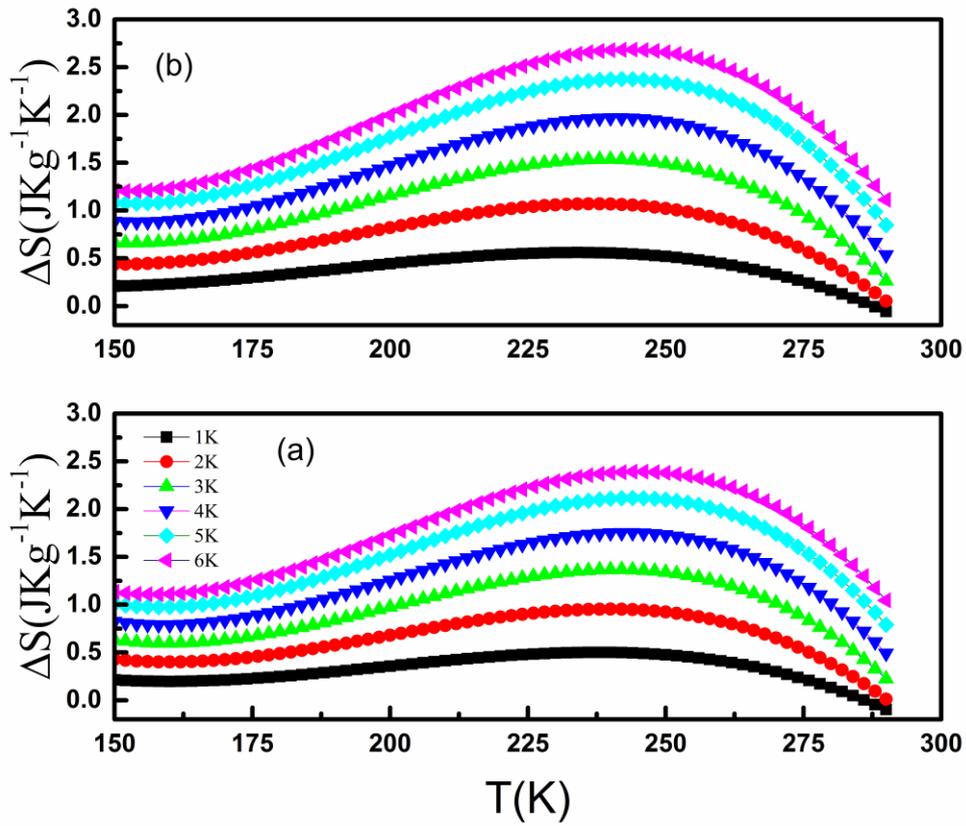

Fig.9 Entropy variation as fonction of temperature in (a) S0 and (b) S5.

The magnetocaloric adiabatic temperature change was calculated in both compounds. The results are plotted in Fig.10. The calculated value of ΔT was slightly higher in the S0 sample with respect to S5 sample. This result is related to the density of spin and spin reodering in S0 in competition with thermal agitation and phonon entropy.

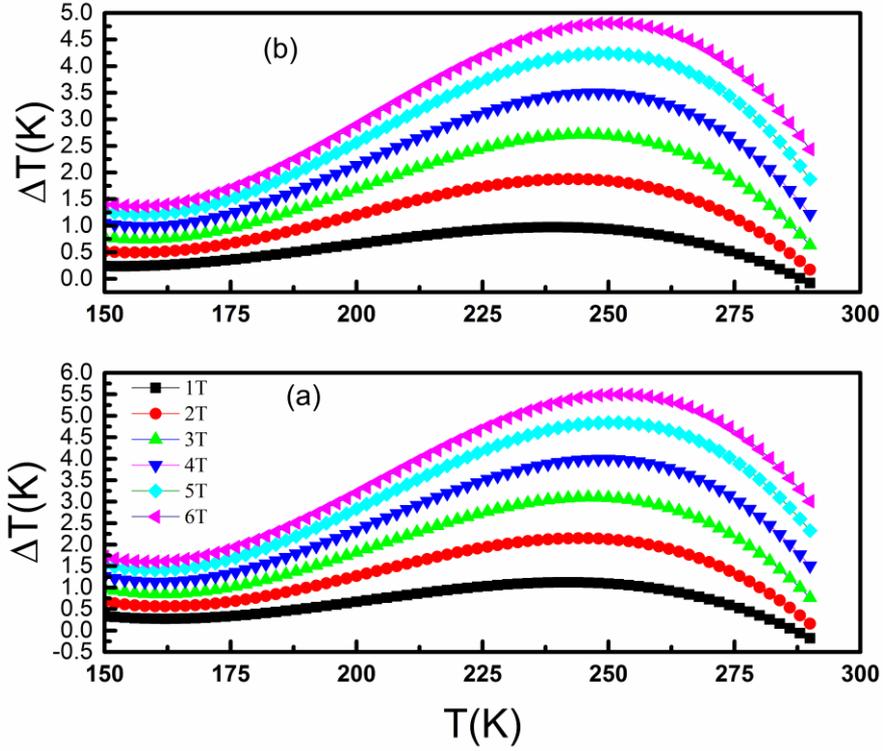

Fig. 10 The magnetocaloric adiabatique temperature variation calculated in (a) S0 and (b) S5.

The examination of the efficiency of our samples in the field of magnetic refrigeration can be done by the calculation of the values of the relative cooling power (RCP) defined by the relation $RCP = -\Delta S_M^{max} \times \delta_{FWHM}$. Here $\Delta S_M^{max}$ is the maximum value of the magnetic entropy, and $\delta T_{FWHM} = (T_{hot} - T_{cold})$ the full width at half maximum of the curve $\Delta S_M$ (T). The results are calculated basing on the curves recorded under the magnetic field of 2T generally delivered by permanent magnet materials that are used in magnetic refrigeration applications. In our samples, the values of the calculated RCP are found to be equal to (79.19 J/Kg) for S5 and (66.78 J/Kg) for S0 sample, the RCP value is found to be enhanced by creating the lacuna. With these results, our samples can be considered as well candidates for the magnetic refrigeration field.

### 3.7. AC magnetic suceptibility measurements

To better understand the magnetocaloric results in the S0 and S5 samples, we performed AC magnetic suceptibility measurements. Fig.11 shows the variation of AC magnetic suceptibility for both samples. These curves were recorderd in FC mode under applied field of 500 Oe. Importantly, the evolution of AC magnetic suceptibility in the two samples is different. Indeed, in the S5 sample the AC magnetic suceptibility showed the very high value in ferromagnetic phase compared to that in S0. This is supported by a gradual and uniform evolution of AC magnetic susceptibility curves without any change of slope, followed by a soft rounding shape before a big break down at Curie temperature. That means that in the S5 sample, spins turne over smoothly when approaching Curie temperature, before a dramatic break down when the temperature went to paraelectric phase. In contrast, in the S0 sample the AC magnetic susceptibility is weak compared to S5. However, when approaching phase transition temperature, from side of ferromagnetic phase, an additional unusial crest is observed in the AC magnetic susceptibility curve in the S0 sample. This crest can be attributed to the pinning of ferromagnetic domains near the curie temperature in the sample S0. This domain pinning is responsible for the amplification of the electrocaloric effect in the S0 sample. We can conclude that the surcose combution method used for sample elaboration produced pinning of ferromagnetic domains in the S0 sample, which is characterised by the observed crest, benefical to magnetocaloric effect (see inset of Fig.11). The Ca-vacancy amplify the AC magnetic suceptibility in this system.

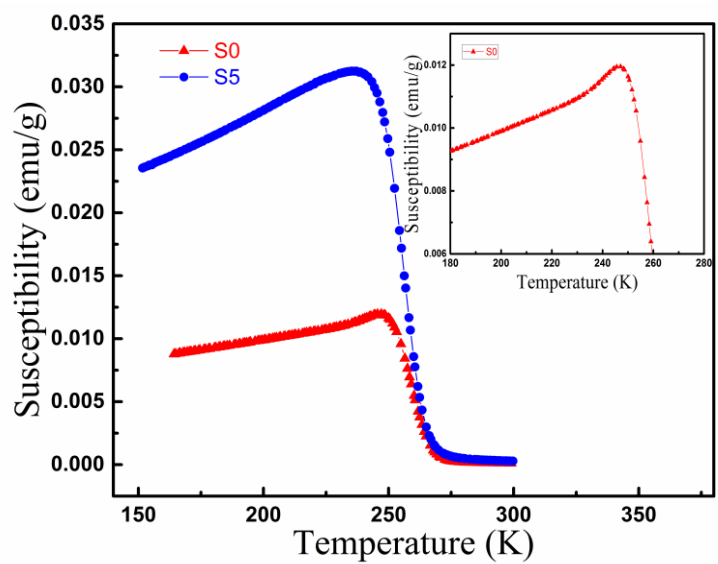

Fig. 11 AC magnetic susceptibility measured in FC mode for S0 and S5. The inset shows the crest appeared in S0 due to elaboration technique.

## 4. Conclusion

Perovskite manganites $La_{0.5}Ca_{0.5-x}\square_x MnO_3$ (LCMO) with x=0 (S0) and x=0.05 (S5) nanomaterials were elaborated by sucrose assisted auto combustion method. The crystalline structure was refined baseing on Rietveld method as the *Pbnm* space group. Paramagnetic-to-ferromagnetic phase transition was observed at 256 K in the S0 sample and at 254 K in the S5 sample. The maximum of entropy change obtained under applied magnetic field of 6T was 2.30 Jkg$^{-1}$K$^{-1}$ and 2.70 Jkg$^{-1}$K$^{-1}$, for S0 and S5 samples respectively. The magnetocaloric adiabatic temperature change value calculated by the indirect method was 5.6 K and 5.2 K for the S0 and S5 samples, respectively. The elaboration method led to the magnetic domains pinning characterized by the appearance of a crest on the AC magnetic susceptibility curve for S0 sample, benifical to the electrocaloric effect. Although, the magnetization and the magnetocaloric adiabatic temperature change are found to be dropped in the Ca-vacancy sample, the AC magnetic susceptibility is highly amplified in this sample. These materials show an improvement of properties by using the autocombustion technique, by creating the lacuna in the A site it has been possible to achieve an increase in the MCE near room temperature. Thus, our samples can be considered as potential manganite perovskite candidates for magnetic refrigeration applications as they are also less costly and environmental friendly, in comparison to Gd-based compounds.

## Acknowledgements

The authors gratefully acknowledge the financial support of the European H2020-MSCA-RISE-2017-ENGIMA action and the CNRST Priority Program PPR 15/2015.